\documentclass[pra,nofootinbib,reprint,pdftex]{revtex4-1}
\usepackage{amsmath,amssymb,amsthm}
\usepackage{graphicx}
\usepackage[pdftex,colorlinks]{hyperref}

\newcommand{\bra}[1]{\left\langle #1 \right|}
\newcommand{\ket}[1]{\left| #1 \right\rangle}

\newcommand{\noisecorrelation}[0]{\alpha}
\newcommand{\noisecoefficient}[0]{A}
\begin{document}

\title{Decoherence strength of multiple non-Markovian environments}
\author{C. H. Fleming}
\affiliation{Joint Quantum Institute and Department of Physics, University of Maryland, College Park, Maryland 20742}
\author{B. L. Hu}
\affiliation{Joint Quantum Institute and Department of Physics, University of Maryland, College Park, Maryland 20742}
\author{Albert Roura}
\affiliation{Max-Planck-Institut f\"ur Gravitationsphysik (Albert-Einstein-Institut), Am M\"uhlenberg 1, 14476 Golm, Germany}
\date{May 21, 2011}

\begin{abstract}
It is known that one can characterize the decoherence strength of a Markovian environment by the product of its temperature and induced damping, and order the decoherence strength of multiple environments by this quantity.
We show that for non-Markovian environments in the weak coupling regime there also exists a natural (albeit partial) ordering of environment-induced irreversibility within a perturbative treatment.
This measure can be applied to both low-temperature and non-equilibrium environments.
\end{abstract}

\maketitle


\section{Introduction}

Environment-induced decoherence is an essential process for a quantum system to acquire classical attributes \cite{Giulini96,Paz99l,Zurek03}.
To characterize how strong an environment can induce decoherence in an open quantum system it is desirable to come up with a measure of the \emph{``decoherence strength''} of each environment acting on the system.
For a Markovian environment, which for quantum Brownian motion refers to high temperatures and ohmic spectral density, such a measure can be constructed by the product of its temperature and damping rate.
However, such a measure may not exist for a \textit{general} environment (with non-ohmic spectral density functions and under low-temperature conditions; see \cite{HPZ92,HPZ93,QBM}) or, even more challenging, for nonequilibrium environments, where the notion of temperature loses meaning.

We show in this paper that at least perturbatively for weak coupling between the system and these environments, low-temperature and non-equilibrium environments \emph{can} be partially ordered.
The object of comparison is, in fact, the correlation function of the collective coupling operators of the environment.
As with matrices and kernels, quantum correlations can only be partially ordered.
However, this does not rule out nontrivial comparisons: in Sec.~\ref{sec:ThermalE} we explicitly detail how one can compare decoherence strengths for combinations of thermal reservoirs without resorting to the concocted notion of an effective temperature, since in general it does not exist.
Our general relation includes some recently reported results \cite{Beer10} as special cases.

It should be noted that strictly speaking it is the ordering of environment-induced irreversibility (in the sense of contraction of the system's state-space volume under time evolution) which will be established here. This involves the phenomena of decoherence, typically dominated by information flow to the environment, as well as thermalization (or equilibration), where energy flow also plays a crucial role. There are many interesting situations (especially when the system-environment coupling is not very strong) in which both phenomena correspond to rather different timescales and one can focus on decoherence as the source of irreversibility at short times. 

Throughout the paper we use units with $\hbar = 1$.

\begin{figure}
\begin{center}
\includegraphics[width=0.45\textwidth]{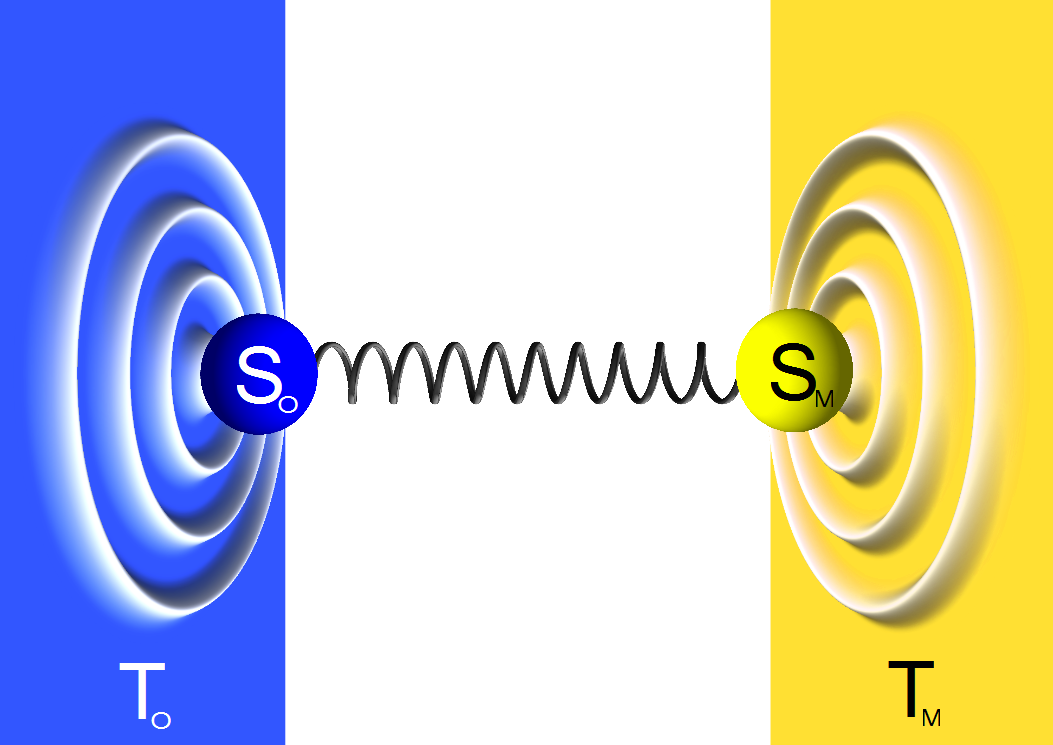}
\end{center}
\caption{\label{fig}Consider, for instance, a system of nano-mechanical resonantors $S_\mathrm{M}$ interacting with a system of optical modes $S_\mathrm{O}$
\cite{Metzger04,Carmon05,Thompson08},
wherein the optical modes experience dissipation and thermal noise $T_\mathrm{O}$ from the cavity field yet the resonators experience dissipation and thermal noise $T_\mathrm{M}$ from a phonon environment.
The combined environment is out of equilibrium and cannot be described by a single spectral-density function and temperature since it does not obey the fluctuation-dissipation relation.
Therefore, the decoherence of this multipartite system cannot be characterized in terms of a temperature and a dissipation coefficient.}
\end{figure}

\section{Quantum correlations \& decoherence strength}
\label{sec:Noise}
Consider a quantum system weakly interacting with an environment and with the interaction Hamiltonian expanded as a sum of separable operators:
\begin{equation}
\mathbf{H}_\mathrm{I} = \sum_n \mathbf{L}_n \otimes \mathbf{l}_n \, , \label{eq:Hint}
\end{equation}
where $\mathbf{L}_n$ and $\mathbf{l}_n$ are system and environment operators respectively.
[The interaction Hamiltonian as well as the set of operators $\mathbf{L}_n$ and $\mathbf{l}_n$ in Eq.~\eqref{eq:Hint} are in the Schr\"odinger picture, but they can in general be time dependent.]

The environment coupling operators $\mathbf{l}_n$ will typically be collective observables of the environment, with dependence upon very many modes.
The system-environment interaction will be treated perturbatively, so that the central ingredient is the (multivariate) correlation function of the environment:
\begin{equation}
\noisecorrelation_{nm}(t,\tau) = \left\langle \underline{\mathbf{l}}_n\!(t) \, \underline{\mathbf{l}}_m\!(\tau) \right\rangle_\mathrm{E} \, , \label{eq:alpha}
\end{equation}
where $\underline{\mathbf{l}}_n\!(t)$ represents the time-evolving $\mathbf{l}_n$ in the interaction (Dirac) picture.
In the \emph{influence functional} formalism \cite{Feynman63} for the quantum Brownian model with bilinear couplings between the system and its environment \cite{CaldeiraLeggett83,HPZ92,HPZ93} the correlation function appears as the kernel in the exponent of a Gaussian influence functional, 
called the influence kernel $\zeta$ in Refs.~\cite{Raval96,Raval97}.
Alternatively, in \emph{quantum state diffusion} \cite{Strunz96} this kernel takes the explicit role of a correlator for complex Gaussian noise.
The influence kernel, or equivalently, the complex correlation function, can be written as a sum of two real parts corresponding to the noise and dissipation kernels \cite{Raval96,Raval97}:
\begin{equation}
\underbrace{\boldsymbol{\noisecorrelation}(t,\tau)}_\mathrm{complex \, noise} = \underbrace{\boldsymbol{\nu}(t,\tau)}_\mathrm{noise} + \, \imath\! \underbrace{\boldsymbol{\mu}(t,\tau)}_\mathrm{dissipation} \, . \label{eq:3kernels}
\end{equation}
The noise kernel $\boldsymbol{\nu}$ appears within the influence functional formalism as the 
correlation of an ordinary real stochastic source,
whereas the dissipation kernel $\boldsymbol{\mu}$ alone would produce a 
purely homogeneous (though not necessarily positivity preserving) evolution.
These same roles can also be inferred from the Heisenberg equations of motion for the system operators after integrating the environment dynamics,
producing the so-called \emph{quantum Langevin equation} \cite{FordOconnell88}.

For weak coupling one can treat perturbatively the system-environment
interaction. Using the notation of Ref.~\cite{QOS}, the second-order master equation \cite{Kampen97,Breuer02a,Strunz04}
for the reduced density matrix $\boldsymbol{\rho}$ of the system can be represented in terms of the noise correlation as
\begin{equation}
\dot{\boldsymbol{\rho}} = \left[ -\imath \mathbf{H}, \boldsymbol{\rho} \right] + \boldsymbol{\mathcal{L}}_2 \{ \boldsymbol{\rho} \} \, ,
\end{equation}
with the second-order contribution given by
\begin{equation}
\boldsymbol{\mathcal{L}}_2 \{ \boldsymbol{\rho} \} \equiv \sum_{nm} \left[ \mathbf{L}_n, \boldsymbol{\rho} \, (\mathbf{\noisecoefficient}_{nm}\! \diamond \mathbf{L}_m)^\dagger - (\mathbf{\noisecoefficient}_{nm}\! \diamond \mathbf{L}_m) \, \boldsymbol{\rho} \right] \, , \label{eq:WCGME}
\end{equation}
where the $\mathbf{\noisecoefficient}$ operators and $\diamond$ product define the second-order operators
\begin{equation}
(\mathbf{\noisecoefficient}_{nm}\! \diamond \mathbf{L}_m)(t) \equiv \int_0^t \!\! d\tau \, \noisecorrelation_{nm}(t,\tau) \, \left\{ \boldsymbol{\mathcal{G}}_0(t,\tau) \, \mathbf{L}_m(\tau) \right\} \, , \label{eq:WCOG}
\end{equation}
with 
$\mathbf{L}_m(\tau)$ in the Schr\"odinger picture and the superoperator $\boldsymbol{\mathcal{G}}_0(t,\tau)$ being the free system propagator, which for a time-independent Hamiltonian simply corresponds to $\boldsymbol{\mathcal{G}}_0(t,\tau) \boldsymbol{\rho}(\tau) = e^{-\imath (t-\tau)\mathbf{H}} \boldsymbol{\rho}(\tau) e^{\imath (t-\tau)\mathbf{H}}$.
Because of the nonlocal character of the noise correlation, the \emph{time-translation generator} in Eq.~\eqref{eq:WCGME} is not generally of Lindblad form.
The theorem by Lindblad \cite{Lindblad76} and Gorini, Kossakowski and Sudarshan (GKS) \cite{Gorini76} specifically characterizes the \emph{algebraic generators} $\boldsymbol{\Phi}$ for all completely-positive maps $e^{\eta \, \boldsymbol{\Phi}}$ ($\eta > 0$ parameterizes the semi-group \cite{Lindblad76,Gorini76} in the Markovian case and will be taken equal to one here) as
\begin{equation}
\boldsymbol{\mathbf{\Phi}} \, \boldsymbol{\rho} =
\underbrace{-\imath \left[ \boldsymbol{\Theta} , \boldsymbol{\rho} \right]}_\mathrm{unitary}
+ \underbrace{\sum_{I,J} \Delta_{IJ} \! \left( \mathbf{e}_I \, \boldsymbol{\rho} \, \mathbf{e}_J^\dagger -\frac{1}{2} \left\{ \mathbf{e}_J^\dagger \mathbf{e}_I , \boldsymbol{\rho} \right\} \right) }_\mathrm{decoherent} \, ,
\label{eq:LindbladFormPhi}
\end{equation}
where $\mathbf{e}_I = \ket{i}\!\bra{i'}$ for some basis $\{ \ket{i} \}$ of the system's Hilbert space and with $I \equiv i\, i'$ 
labeling all possible pairs of indices.
Such generators and the dynamics they engender when the master equation has the Lindblad form have been extensively studied \cite{Kossakowski72,Davies74,Davies76a,Davies77,Alicki07,Accardi02,Attal06,Ingarden97,Lindblad83,Weiss93}.
The (algebraic) \emph{dissipator} $\Delta_{IJ}$ is a positive-definite 
matrix%
\footnote{The master equation for $\dot{\boldsymbol{\rho}}$ exhibits the same structure as the right-hand side of Eq.~\eqref{eq:LindbladFormPhi}. However, whereas in the Markovian regime one would simply have in place of $\boldsymbol{\Theta}$ and $\Delta_{IJ}$ their time derivatives, this is no longer the case in the non-Markovian regime because $\dot{\boldsymbol{\mathbf{\Phi}}}$ and $\boldsymbol{\mathbf{\Phi}}$ do not commute.}.
Note that the ``dissipation'' generated by the dissipator is that of states, including decoherence. In fact, this notion of ``state dissipation'' can be given a more precise geometrical meaning as follows.
For any distance $D$ on the space of density operators which is constructed from a \emph{monotonic} metric (e.g.\ trace distance or Bures distance),
completely positive evolution cannot cause any Hilbert-space distances to expand \cite{Petz96}:
\begin{equation}
D \big[ \boldsymbol{\mathcal{G}}(t) \boldsymbol{\rho}_1 ,
\boldsymbol{\mathcal{G}}(t) \boldsymbol{\rho}_2 \big]
\leq D [ \boldsymbol{\rho}_1 , \boldsymbol{\rho}_2 ] \, .
\end{equation}
From this result it is then easy to prove that positive-definite dissipators contract the state-space volume,
whereas negative-definite dissipators expand the state-space volume (since they appear to be time-reversed contractions).

Derived in App.~\ref{sec:Lindblad} and also with more context in Ref.~\cite{QOS}, the second-order algebraic dissipator $\Delta_{IJ}(t)$ evaluates to
\begin{equation}
\sum_{nm} \int_0^t \!\! d\tau \!\int_0^t \!\! d\tau' \bra{i} \underline{\mathbf{L}}_m(\tau) \ket{i'} \noisecorrelation_{nm}(\tau'\!,\tau) \overline{\bra{j} \underline{\mathbf{L}}_n(\tau') \ket{j'}} \, , \label{eq:LindbladAG}
\end{equation}
in the interaction (Dirac) picture. 
Expression \eqref{eq:LindbladAG} will be shown to be a positive-definite quadratic form for all microscopically derived noise correlations, thus agreeing with the Lindblad-GKS theorem in as much as is required.
Although the Markovian generators which appear in the Lindblad equation are well known, to our knowledge these non-Markovian generators (which are strictly algebraic and do not appear in the master equation) are a novel discovery.

The key to accomplishing our stated goal rests in the comparison of perturbative dissipators.
We first note that from its microscopic origins, Eq.~\eqref{eq:alpha}, the 
environment correlation function is Hermitian in the sense of
\begin{equation}
\boldsymbol{\noisecorrelation}(t,\tau) = \boldsymbol{\noisecorrelation}^{\!\dagger}\!(\tau,t) \, , \label{eq:alpha_her}
\end{equation}
where the boldface notation denotes a matrix with respect to the indices $n,m$ in Eq.~\eqref{eq:alpha}, 
and also positive definite in the sense of
\begin{equation}
\int_0^t \!\! d\tau_1 \! \int_0^t \!\! d\tau_2 \, \mathbf{f}^\dagger(\tau_1) \, \boldsymbol{\noisecorrelation}(\tau_1,\tau_2) \, \mathbf{f}(\tau_2) \geq 0 \, , \label{eq:posdef1}
\end{equation}
for all vector functions $\mathbf{f}(t)$ indexed by the environment correlator. 
All quantum correlations lead at least to \emph{accumulated decoherence} since their algebraic Lindblad dissipator, Eq.~\eqref{eq:LindbladAG}, is necessarily positive definite.
Accumulated decoherence only implies that there is more net decoherent evolution than recoherent evolution.
In general the stricter property of instantaneous decoherent evolution, $\dot{\boldsymbol{\Delta}}(t) > 0$ (boldface denotes here a matrix with indices $I,J$),
can only be satisfied by environements with local correlation function (Markovian processes) and would always produce a Lindblad master equation.
However, some very restricted classes of system-environment interactions, such as the interaction Hamiltonian in the rotating-wave approximation (RWA) \cite{RWA}, can be constrained by the particular form of their coupling to be instantaneously decoherent.
This characterizes the class of systems with non-Markovian dynamics (nonlocal environment correlation function) whose master equation is, nevertheless, naturally of Lindblad form, though not necessarily at all times.

The key result of this work is showing that the 
environment correlation function itself provides a natural comparison of state dissipation or \emph{decoherence strength}.
If two correlation functions are ordered $\boldsymbol{\noisecorrelation}_+(t,\tau) > \boldsymbol{\noisecorrelation}_-(t,\tau)$ in the sense of positivity~\eqref{eq:posdef1},
then their corresponding second-order Lindblad dissipators are also ordered $\boldsymbol{\Delta}_+(t) > \boldsymbol{\Delta}_-(t)$,
and we can, therefore, say that one environment generates more state dissipation than the other, regardless of the system.
For instance, the set of univariate Markov processes is totally ordered by the scalar magnitude of the respective delta correlations,
e.g. $2 \, \delta(t-\tau) > 1 \, \delta(t-\tau)$.
In general, the set of all quantum correlations is only partially ordered, but nontrivial
orderings do exist. We illustrate this principle with several examples below.

\section{Thermal correlations} \label{sec:ThermalE}

\subsection{Individual reservoirs}

Time-independent coupling to a thermal reservoir will always produce time-translation-invariant environment correlations which can be expressed in the Fourier domain as
\begin{eqnarray}
\tilde{\boldsymbol{\noisecorrelation}}(\omega) &=& \tilde{\boldsymbol{\gamma}}(\omega) \left[ \tilde{\kappa}_T(\omega) - \omega \right] \, , \label{eq:alphaThermal} \\
\tilde{\kappa}_T(\omega) & \equiv & \omega \, \coth\!\left( \frac{\omega}{2T} \right) \, ,
\end{eqnarray}
in terms of the damping kernel $\tilde{\boldsymbol{\gamma}}(\omega)$ (anti-derivative of the dissipation kernel) and fluctuation-dissipation kernel $\tilde{\kappa}_T(\omega)$.
This can be derived directly from first principles in Eq.~\eqref{eq:alpha},
by demanding a coupling-invariant fluctuation-dissipation relation (FDR),
or by demanding coupling-invariant detailed balance in the master equation \cite{QOS}.

Note that a Markovian quantum regime (complex white noise) necessarily implies a local damping kernel and high temperature (local FDR kernel).
The Markovian regime is reached when the system time scales are much slower than those of the environment,
so that we can take the zero-frequency approximation
\begin{equation}
\lim_{\omega \to 0} \tilde{\boldsymbol{\noisecorrelation}}(\omega) = \tilde{\boldsymbol{\gamma}}(0) \, 2 T \, .
\end{equation}
Markovian processes can, therefore, be ordered in their decoherence strength by the product of their damping and temperature, a result which is well known.
If one inquires as to the temperature of an unknown Markovian process, the FDR kernel (or noise-to-damping ratio) will always reveal this.

In general, nonlocal correlations (e.g.\ for finite temperature), and thus decoherence strengths, are not totally ordered.
For a fixed temperature thermal correlations can be ordered by damping.
On the other hand we have the inequality
\begin{equation}
\tilde{\kappa}_{\mathrm{hot}}(\omega) > \tilde{\kappa}_{\mathrm{cold}}(\omega) \geq |\omega| \, , \label{eq:hotandcold}
\end{equation}
and so for fixed damping, correlations can also be ordered by their temperature.
Therefore, finite-temperature thermal correlations are partially ordered by damping and temperature.
\begin{equation}
\tilde{\boldsymbol{\gamma}}_\mathrm{strong}(\omega) \left[ \tilde{\kappa}_\mathrm{hot}(\omega) - \omega \right] > \tilde{\boldsymbol{\gamma}}_\mathrm{weak}(\omega) \left[ \tilde{\kappa}_\mathrm{cold}(\omega) - \omega \right] \, ,  \label{eq:EqDec}
\end{equation}
and so it immediately follows that
\begin{equation}
\tilde{\boldsymbol{\noisecorrelation}}_\mathrm{strong}^\mathrm{hot}(\omega) > \tilde{\boldsymbol{\noisecorrelation}}_\mathrm{weak}^\mathrm{cold}(\omega) \, .
\label{eq:ordering2}
\end{equation}
If one environment has weaker damping but a sufficiently higher temperature, then the two correlations cannot be ordered --
the implication is that different systems would decohere faster or slower for each environment, but not in a manner which can be strictly ordered.

From Eq.~\eqref{eq:EqDec} we can now compare environments of low temperature and nonlocal damping.
For fixed damping, the monotonic ordering of temperature is no surprise.
While for fixed temperature, the ordering of damping is more subtle though also not surprising.
The damping can be increased by an overall rescaling, say $\tilde{\boldsymbol{\gamma}}(\omega) \to 2 \, \tilde{\boldsymbol{\gamma}}(\omega)$,
or multiplying it by a frequency-dependent function $g(\omega)$ such that $g(\omega) \geq 1, \ \forall \omega$.

\subsection{Multiple environments}
\label{sec:ETK}
Here we wish to recover and generalize the work of Beer \& Lutz \cite{Beer10} wherein they 
compared the decoherence rates of collective environments with different temperatures and ohmic cutoff frequencies,
specifically for linear coupling to an oscillator and with both reservoirs at a relatively high temperature.
For multiple environments we can make the same comparison by using the natural measure of decoherence strength from the Lindblad dissipator and environment correlation function.
First we note that for coupling to one reservoir, the individual thermal correlations can be expressed via Eq.~\eqref{eq:alphaThermal}.
Next we note that for any monotonic cutoff regulator, with fixed local limit $\tilde{\gamma}(0)$
and variable cutoff $\Lambda$, then
\begin{equation}
\tilde{\boldsymbol{\gamma}}_\mathrm{high}(\omega) \geq \tilde{\boldsymbol{\gamma}}_\mathrm{low}(\omega) \, ,
\end{equation}
where $\Lambda_\mathrm{high} > \Lambda_\mathrm{low}$ (referred to as ``fast'' and ``slow'' in Ref.~\cite{Beer10}).
We can also compare the individual FDR kernels as per Eq.~\eqref{eq:hotandcold}.
Finally we can use the above relations to construct the mathematical inequality
\begin{equation}
[ \tilde{\boldsymbol{\gamma}}_\mathrm{high}(\omega) - \tilde{\boldsymbol{\gamma}}_\mathrm{low}(\omega) ] \{ [\tilde{\kappa}_\mathrm{hot}(\omega) -\omega ] - [ \tilde{\kappa}_\mathrm{cold}(\omega) -\omega ] \}  > 0 \, ,
\label{eq:ordering3}
\end{equation}
which can then be rearranged to show that
\begin{equation}
\tilde{\boldsymbol{\noisecorrelation}}_\mathrm{high}^\mathrm{hot}(\omega) + \tilde{\boldsymbol{\noisecorrelation}}_\mathrm{low}^\mathrm{cold}(\omega) > \tilde{\boldsymbol{\noisecorrelation}}_\mathrm{low}^\mathrm{hot}(\omega) + \tilde{\boldsymbol{\noisecorrelation}}_\mathrm{high}^\mathrm{cold}(\omega) \, , \label{eq:Lutz}
\end{equation}
which is consistent with the results of Ref.~\cite{Beer10}, when we interpret the left and right-hand sides of Eq.~\eqref{eq:Lutz} as comparing the decoherence strengths of two different collective (non-equilibrium) environments.
Note that as the individual reservoirs are Gaussian and independent, one may simply add their correlations in determining the collective correlation.
Our result applies more generally than that of Ref.~\cite{Beer10}, in terms of coupling and temperature, though we do not calculate a specific decoherence time.
Their work relied upon what is essentially the exact FDR kernel, but which has been referred to as an \emph{effective temperature} \cite{Cugliandolo97} in the classical regime.
We would rather avoid this nomenclature given that such environments will  lead to an asymptotic stationary state which is not thermal in general (sufficiently simple systems can reach a thermal state, but the corresponding temperature will be different for different systems).

\section{Discussion}
In this work we have motivated a general notion of decoherence strength as that generated by the quantum correlations of the environment,
which in turn determine the magnitude of the algebraic Lindblad dissipator and thus ``state dissipation'' itself (see Sec.~\ref{sec:Noise} for its precise meaning).
The ordering of decoherence strengths in this formalism is only partial, although when it occurs it is independent of the initial state of the system and of the the particular operators $\mathbf{L}_n$ which characterize the system coupling to the environment.
However, this is not to say that comparison of environment correlations is not useful when there is no strict ordering. 

State-dependent decoherence is of particular interest in the search for decoherence-free states and how they emerge in certain classes of environments.
If for two environments $\tilde{\boldsymbol{\noisecorrelation}}_1(\omega) - \tilde{\boldsymbol{\noisecorrelation}}_2(\omega)$ is indefinite,
then there \emph{could} be states corresponding to the vectors $\mathbf{f}$ in Eq.~\eqref{eq:posdef1} which exploit this.
As an explicit example, the well-known phenomena of sub-radiant and super-radiant spontaneous emission of atoms in the electromagnetic field vacuum can be viewed in this manner.
A detailed analysis can be found in Ref.~\cite{Dipole}, which is formulated most directly in terms of the electromagnetic field correlation function.
A well-known result covered there is that for a pair of two-level atoms very close together there is a joint state which decoheres rapidly (super-radiance) and another joint state which decoheres very slowly (sub-radiance), as compared to the decoherence rate of an isolated atom.
This is possible because when two atoms are brought close together in the electromagnetic field, their multivariate correlation $\tilde{\boldsymbol{\noisecorrelation}}_\mathrm{near}(\omega)$ cannot be totally ordered with respect to the factorized correlations present in the far-distance limit, $\tilde{\boldsymbol{\noisecorrelation}}_\mathrm{far}(\omega)$.

As mentioned in the introduction and defined precisely in Sec.~\ref{sec:Noise},  strictly speaking the ordering that we have established is for environment-induced irreversibility (in terms of ``state dissipation''). For weak coupling, one expects that in most situations of interest decoherence will dominate over thermalization at short times in such an  irreversibility process. In fact, if one considers times much shorter than the relaxation timescale(s) and the real part of the $\boldsymbol{\noisecorrelation}(t,\tau)$ can be neglected, the decoherence strength for an environment in equilibrium is characterized by $\tilde{\boldsymbol{\gamma}}(\omega) \tilde{\kappa}_T(\omega)$. Furthermore, from Eqs.~\eqref{eq:EqDec} and \eqref{eq:ordering3} one can immediately see that the same inequalities as in Eqs.~\eqref{eq:ordering2} and \eqref{eq:Lutz} also hold for $\tilde{\boldsymbol{\gamma}}(\omega) \tilde{\kappa}_T(\omega)$.

\section*{Acknowledgments}
We would like to acknowledge Dr.\ Eric Lutz for drawing our attention to his calculation of decoherence times for multiple environments.
This work is supported in part by NSF grants PHY-0426696, PHY-0801368, DARPA grant DARPAHR0011-09-1-0008 and the Laboratory for Physical Sciences.

\appendix
\section{Non-Markovian dissipators}\label{sec:Lindblad}
As mentioned in Sec.~\ref{sec:Noise}, application of the Lindblad-GKS theorem in the non-Markovian regime requires,
not the time-translation generator, but the algebraic generator.
To obtain the non-Markovian dissipator quickly, we will first begin with the von Neumann equation for the density-matrix propagator of the closed combined system C, consisting of system + environment, in the interaction (Dirac) picture:
\begin{align}
\frac{d}{dt} \underline{\boldsymbol{\mathcal{G}}}_\mathrm{C}(t) &= \underline{\boldsymbol{\mathcal{L}}}_\mathrm{I}(t) \, \underline{\boldsymbol{\mathcal{G}}}_\mathrm{C}(t) \, , \label{eq:SEmaster}
\end{align}
where $\underline{\boldsymbol{\mathcal{G}}}_\mathrm{C}(t)$ is the interaction-picture propagator
and $\underline{\boldsymbol{\mathcal{L}}}_\mathrm{I}(t)$ is the interaction-picture Liouvillian, defined as
\begin{align}
\underline{\boldsymbol{\mathcal{G}}}_\mathrm{C}(t) &= \boldsymbol{\mathcal{G}}_\mathrm{F}^{-1}(t) \, \boldsymbol{\mathcal{G}}_\mathrm{C}(t) \, , \\
\underline{\boldsymbol{\mathcal{L}}}_\mathrm{I}(t) &= \boldsymbol{\mathcal{G}}_\mathrm{F}^{-1}(t) \, \boldsymbol{\mathcal{L}}_\mathrm{I}(t) \, \boldsymbol{\mathcal{G}}_\mathrm{F}(t) \, ,
\end{align}
in terms of the free (non-interacting) system + environment propagator $\boldsymbol{\mathcal{G}}_\mathrm{F}(t)$.
Note that $\underline{\boldsymbol{\mathcal{G}}}_\mathrm{C}(t)$ are superoperators, which act on usual Hilbert-space operators as $\underline{\boldsymbol{\mathcal{G}}}_\mathrm{C}(t) \, \mathbf{A} = \underline{\boldsymbol{\mathcal{U}}}(t,0) \, \mathbf{A} \,\,\underline{\boldsymbol{\mathcal{U}}}^{-1}(t,0)$, and similarly for $\boldsymbol{\mathcal{G}}_\mathrm{F}(t)$ and its inverse.
The interaction Liouvillian is furthermore defined in terms of the interaction Hamiltonian $\mathbf{H}_\mathrm{I}(t)$ as 
\begin{equation}
\boldsymbol{\mathcal{L}}_\mathrm{I}(t) \, \boldsymbol{\rho} = -\imath \left[ \mathbf{H}_\mathrm{I}(t) , \boldsymbol{\rho} \right] \, .
\end{equation}

For simplicity we consider factorized initial states of the system and environment so that we can solve Eq.~\eqref{eq:SEmaster} with a Neumann series and then trace out the environment more easily.
(Properly correlated initial states of the system and environment within this formalism are considered in Ref.~\cite{Correlations}.)
The Neumann series produces a perturbative expansion of the open-system propagator:
\begin{align}
\underline{\boldsymbol{\mathcal{G}}}(t) &
= \mathrm{T} \, \exp\!\left[\int_0^t \! d\tau \, \underline{\boldsymbol{\mathcal{L}}}_\mathrm{I}(\tau) \right]
= 1+ \sum_{k=1}^\infty \underline{\boldsymbol{\mathcal{G}}}_k(t) \, , \\
\underline{\boldsymbol{\mathcal{G}}}_k(t) &= \left\langle \prod_{i=1}^k \int_0^{\tau_{i-1}} \!\! d\tau_{i} \, \underline{\boldsymbol{\mathcal{L}}}_\mathrm{I}(\tau_{i}) \right\rangle_{\!\!\!\mathrm{E}} \, ,
\end{align}
also in the interaction picture and where $\tau_0=t$.
This series can be contracted into the single exponential
\begin{eqnarray}
\underline{\boldsymbol{\mathcal{G}}}(t) &=& e^{\underline{\boldsymbol{\Phi}}(t)} \, , \\
\boldsymbol{\mathcal{G}}(t) &=& \boldsymbol{\mathcal{G}}_0(t) \, e^{\underline{\boldsymbol{\Phi}}(t)} \, ,
\end{eqnarray}
where $\boldsymbol{\mathcal{G}}_0(t)$ is the free propagator for the system.
For symmetric noise (with vanishing odd cumulants), such as Gaussian with vanishing mean, the perturbative generators can then be found to be 
\begin{eqnarray}
\underline{\boldsymbol{\Phi}}_{2}(t) &=& \underline{\boldsymbol{\mathcal{G}}}_2(t) \, , \\
\underline{\boldsymbol{\Phi}}_{4}(t) &=& \underline{\boldsymbol{\mathcal{G}}}_4(t) - \frac{1}{2} \underline{\boldsymbol{\mathcal{G}}}_2^2(t) \, .
\end{eqnarray}
This is equivalent to solving the master equation via \emph{Magnus series}~\cite{Magnus54} in the interaction picture.
It should be noted that Magnus-series solutions are slightly secular in time, since in general the Magnus series has a finite radius of convergence \cite{Blanes98}.
In this context the second-order Magnus-series solution will accurately match the correct solution to second order at early times, and then only match the correct solution to zeroth order at later times, wherein it converges to the RWA solution.
For some aspects of the solution this accuracy can be improved with a less-secular integrator.
However, a careful analysis of the master equation and its solutions has shown that, due to unavoidable degeneracy, the second-order master equation is fundamentally incapable of providing the full second-order solutions \cite{Accuracy}.
Therefore we would not promote these solutions as \emph{the} second-order solutions,
but they contain the most information pertaining to the non-Markovian dissipation, which one can extract from the second-order master equation.

The Magnus-series solution to the second-order master equation gives rise to the second-order algebraic generator
\begin{equation}
\underline{\boldsymbol{\Phi}}(t) = \int_0^t \!\! d\tau \! \int_0^\tau \!\! d\tau' \left\langle \underline{\boldsymbol{\mathcal{L}}}_\mathrm{I}(\tau) \, \underline{\boldsymbol{\mathcal{L}}}_\mathrm{I}(\tau') \right\rangle_{\!\mathrm{E}} + \mathcal{O}(\boldsymbol{\mathcal{L}}_\mathrm{I}^4) \, ,
\end{equation}
in the interaction picture.
In terms of the interaction Hamiltonian, the Lindblad coefficients of the algebraic generator $\underline{\boldsymbol{\Phi}}(t)$ are then
\begin{equation}
\underline{\Delta}_{IJ}(t) = \int_0^t \!\! d\tau \! \int_0^t \!\! d\tau' \bigl\langle \bra{i} \underline{\mathbf{H}}_\mathrm{I}(\tau) \ket{i'} \bra{j'} \underline{\mathbf{H}}_\mathrm{I}(\tau') \ket{j} \bigr\rangle_{\!\mathrm{E}} \, ,
\end{equation}
given the representation in Eq.~\eqref{eq:LindbladFormPhi}.
With the interaction Hamiltonian expanded as a sum of separable operators, as per Eq.~\eqref{eq:Hint}, the coefficients evaluate to
\begin{equation}
\sum_{n,m} \int_0^t \!\! d\tau \!\int_0^t \!\! d\tau' \bra{i} \underline{\mathbf{L}}_m(\tau) \ket{i'} \alpha_{nm}(\tau'\!,\tau) \, \overline{\bra{j} \underline{\mathbf{L}}_n(\tau') \ket{j'}} \, ,
\end{equation}
in terms of the environment correlation function.
Both forms are positive-definite quadratic forms.
This implies that the second-order master equation must generate completely-positive maps to second order. Furthermore, it also implies that
the second-order Magnus-series solution $\boldsymbol{\mathcal{G}}(t) = \boldsymbol{\mathcal{G}}_0(t) \, e^{\underline{\boldsymbol{\Phi}}_2\!(t)}$ is actually completely positive \emph{exactly} (rather than just through second order).

Therefore, one can see that the correlation function $\boldsymbol{\alpha}(t,\tau)$ is not only the influence kernel and complex noise kernel, but also a decoherence kernel which determines the magnitude of the non-Markovian dissipator, at least in the weak-coupling limit.
Note that in this derivation the correlation function may describe multivariate noise arising from an environment completely out of equilibrium.

\bibliography{bib}{}
\bibliographystyle{apsrev4-1}
\end{document}